\def\lae{\mathrel{<\kern-1.0em\lower0.9ex\hbox{$\sim$}}}
\def\etal {et~al.}

\documentclass[12pt,pp]{aastex}

\newcommand{\Msun}{\mbox{ M}_{\odot}}
\newcommand{\gae}{\mathrel{>\kern-1.0em\lower0.9ex\hbox{$\sim$}}}

\newcommand{\NV}{\ion{N}{5}\,$\lambda\lambda\,1238.8,\,1242.8$\ }
 
\newcommand{\SiIV}{\ion{Si}{4}\,$\lambda\lambda\,1393.7,1402.8$\ }

\lefthead{Boroson and Vrtilek}

\begin{document}

\title{X-ray Variations at the Orbital Period from Cygnus X-1 in the
High/Soft State}

\author{Bram Boroson}
\affil{Smithsonian Astrophysical Observatory
Mail Stop 83
Cambridge, MA 02138
bboroson@cfa.harvard.edu}

\and

\author{Saeqa Dil Vrtilek}
\affil{Smithsonian Astrophysical Observatory
Mail Stop 67
Cambridge, MA 02138
svrtilek@cfa.harvard.edu}


\begin{abstract}
Orbital variability has been found in the X-ray hardness of the
black hole candidate Cygnus X-1 during the soft/high X-ray state using 
light curves provided by the Rossi X-ray Timing Explorer's All Sky
Monitor. We
are able to set broad limits on how the mass-loss rate and
X-ray luminosity vary between the hard and soft states.
The folded light curve shows diminished flux in the
soft X-ray band at $\phi=0$ (defined as the time of of the
superior conjunction of the X-ray source). 
Models of the orbital variability provide slightly superior fits when
the absorbing gas is concentrated in neutral clumps and better explain
the strong variability in hardness.
In combination with the previously established
hard/low state dips, our observations 
give a lower limit to the mass loss rate in the soft state
($\dot{M}<2\times10^{-6}\Msun$~yr$^{-1}$) than the limit in
the hard state ($\dot{M}<4\times10^{-6}\Msun$~yr${-1}$). 
Without a change in the wind structure between X-ray states, the greater
mass-loss rate during the low/hard state would be inconsistent with the
increased flaring seen during the high-soft state.
\end{abstract}


\section{Introduction}

X-ray binary systems in which the compact object is thought to be a
black hole from dynamical or X-ray timing evidence (Black Hole
Candidates, or BHC, see Remillard \&\ McClintock 2006) are often found
to shift between spectral states distinguished by hardness and intensity
at X-rays and $\gamma$-rays (Grove et al. 1998). This class of source
comprises several objects of which the BHC Cygnus~X-1 is a
member (Grebenev et al. 1993). 

Cygnus X-1 is unusual in that it is persistently bright in X-rays and has a 
high mass donor star, HDE~226868, with mass 18-40~M$_\odot$ (Zi\'olkowski 
2005; Tarasov et al. 2003; Caballero-Nieves et al. 2009) and accretes 
primarily from a stellar wind. In contrast most BHC are transients (hence 
cannot be revisited in coordinated observing campaigns) and have low mass 
companions that accrete via Roche lobe overflow.  The compact object in 
Cygnus~X-1 has a mass of 7-20~M$_\odot$ (Herrero et al. 1995; Zi\'olkowski 
2005). No pulsations or bursts have ever been detected.  The strong stellar 
wind of the donor star, and its complex interaction with the compact 
object, can be studied through the UV P-cygni lines and X-ray absorption 
(refs).

The binary orbital period of $P=5.6$ days was first discoved by Bolton 
(1972) and refined by Brocksopp~\etal~(1999b) using radial velocities of 
the O-star. Brocksopp~\etal~ found P = 5.599829$\pm$0.000016d, with an 
epoch of the inferior conjuction of the O-star of To=41874.207$\pm$0.009MJD 
and we use their values. The X-ray phases computed from this ephemeris are 
accurate to 0.007 in phase through the end of the ASM data set, which is 
sufficient for our purposes.

Gies~et al.~(2003) used the results of a four year spectroscopic
monitoring of the H$\alpha$ emission strength of HDE226868, the noncompact
companion to Cyg X-1, to conclude that the changes in X-ray state are
driven by changes in $\dot{M}$, the mass-loss rate through the wind.
Gies~et al. find that $\dot{M}$ is lower during the soft/high state
and suggest that in this
state, the X-rays ionize the wind to a greater extent and slow its
acceleration. As a result, the black hole can accrete more matter and
the total X-ray luminosity increases.

Spectra obtained with the {\it Chandra} High Energy Transmission
Gratings suggest that the stellar wind increases in density towards
the
compact object. This ``focused wind'' may cause increased X-ray
absorption at orbital phases from $\phi=0.7-0.9$. Miller et al.~(2005)
found that at
$\phi=0.76$ the X-ray shows strong absorption lines, consistent with
this scenario. Further observations of X-ray spectra during this phase
interval have returned varying results.
Schulz~et
al.~(2002) suggest some X-ray lines may have P~Cygni profiles at
$\phi=0.93$, although Marshall~et al.~(2001) did not find such
features
at $\phi=0.84$. Hanke et al. (2009)
pointed out that the previous
observations of the focused wind were made at times when the RXTE ASM
count rate varied by nearly a factor of~3.

The behavior of lines in the UV does not support the focused wind
picture.
The wind-formed UV P~Cygni absorption gradually decreases as the
compact object
moves towards the line of sight, whereas a focused wind might be
expected to cause a sharp increase in the absorption (Vrtilek et
al. 2008).
The general trend of the P~Cygni line variation is consistent with
the Hatchett-McCray effect (Hatchett \&\ McCray 1977), in which an orbiting
X-ray source removes ions from a stellar wind. The UV spectrum
has been observed with {\it IUE} (Dupree et al. 1978, Treves et al.
1980, van Loon et al. 2001), which has shown that absorption in the
P~Cygni lines of N\,{\sc V}, C\,{\sc IV}, and Si\,{\sc IV} diminishes
near $\phi=0.5$. Recent analyses of Hubble Space Telescope Imaging
Spectrograph (HST STIS) observations of Cyg~X-1 during the soft/high
state provided more detailed diagnostics of the structure of the stellar
wind (Vrtilek et al. 2008, Geis et al. 2008).

The X-ray light curve of Cyg~X-1 shows intensity dips near $\phi=0$. This was first
observed with Copernicus (Mason et al. 1974, Treves et al. 1980) and
Ariel 5 (Holt et al. 1979). Power spectra of a 100~day time series
collected with WATCH
(Priedhorsky, Brandt, \&\ Lund 1995) showed a dip of $\approx20$\% near
$\phi=0$. In an analysis of 2 years of RXTE/ASM data of Cyg~X-1,
Wen~\etal~(1999) found variability at the 5.6 day orbital period during
the X-ray low/hard state, but no evidence of orbital periodicity during
the high/soft state. Similar results were found by Brocksopp et al.
(1999a), who also studied simultaneous orbital variability in optical,
IR, and radio light curves. Absorption of X-rays by a stellar wind from
the companion star can reproduce the observed X-ray orbital modulations
in the hard state. However, the RXTE ASM observations covered
$\approx900$~days and only one soft/high state at the time of the Wen et
al. analysis. 
As the RXTE ASM data have accumulated, the Cyg~X-1 data set
has been revisited for analysis. Baluci\'{n}ska-Church et al. (2000)
identified discrete dipping events from a number of X-ray observatories
including the RXTE ASM and showed that their distribution in the
hard/low state peaks near $\phi=0$ with a secondary peak near
$\phi=0.6$.  Pointed observations with RXTE over the course of Cyg~X-1's
orbit suggest that there are at least two different kinds of
X-ray dips in Cyg~X-1 (Feng \&\ Cui 2002). Dips can either
preferentially affect the low-energy spectrum or can decrease the
spectrum in a nearly energy-independent manner.

Lachowicz et
al. (2006) and Ibragimov, Zdziarski, \&\ Poutanen (2007) found evidence
in the RXTE ASM data for a superorbital period and a beat period between
the orbital and superorbital periods. Lachowicz et al. (2006) looked for
but did not find evidence for orbital periodicity in the soft/high
period. Simultaneous radio observations did, however, show the 5.6~d
orbital period. Dong, Wang, \&\ Xue (2007) reported evidence for
variability at the 5.6~d period during the soft state, in addition to
periodicities at 1.0$\pm0.2$~d and 18.0$\pm3.0$~d. They did not evaluate
the significance of these periodicities, however, nor did they examine
the fractional variability or folded light curves of these variations.

The RXTE~ASM data set now covers more than 4000~days and about 5
soft/high states. A larger data set should be more sensitive to weak
variability and may display states which the source enters into less
frequently. 

Here we show convincing evidence for the 5.6~d period during the soft
state by using the more extensive data set accumulated by RXTE and
by examining the variability of the X-ray colors, that is, the ratio of
X-ray flux between different energy bands. We examine the nature of this
orbital variability and use the observations to set limits on how the
stellar wind may change between the soft and hard states.

\section{Observations}

%

The Rossi X-ray Timing Explorer (RXTE) (Bradt, Rothschild, \&\ Swank
1993), launched in 1995, includes an All Sky Monitor (ASM, Levine et
al. 1996). The ASM contains three scanning shadow cameras (SSCs) with
total effective area of 90~cm$^2$. 

In addition to the ASM, RXTE contains a Proportional Counter Array
(PCA, Jahoda et al. 1996) and High Energy X-ray Timing Experiment
(HEXTE, Rothschild et al. 1998). The PCA has an effective area
of $\sim 6500$~cm$^{2}$ and HEXTE has an effective area of
$\sim1000$~cm$^{-2}$, allowing spectra in the 2--200~keV range
to be obtained with a variety of timing modes down to
$\sim1\mu$s.

\subsection{RXTE ASM}

The RXTE ASM Products Database provides light curves and intensities in
the 1.5-3, 3-5, and 5-12 keV bands (``colors") for $\approx350$ sources.
We use the Definitive Products version of the data that has been
processed at MIT. 
In the resulting time series, there are often two measurements that
occur at exactly the same time.  This is because two of the SSCs have
overlapping fields of view and can provide independent measurements. For
our analysis, when this occurs we average the two
measurements. For the ASM observatons of Cyg~X-1, 70\%\ of the
measurements carry a unique time stamp.
The error bars on the light curves and colors are based on counting
statistics convolved with a 3\% systematic error, as estimated by the
RXTE ASM team from Crab pulsar measurements.

\begin{deluxetable}{lrrr}
\tablecolumns{4}
\tablewidth{0pc}
\tablecaption{Starting and ending MJD defining the Cyg X-1 soft states
referred to in the text.}
\tablehead{
\colhead{Label} & \colhead{Start (MJD)} & \colhead{Stop (MJD)} &
\colhead{Duration (d)}\\}
\startdata
Soft 1 & 50,220 & 50,300 & 200\\
Soft 2 & 51,850 & 51,950 & 50\\
Soft 3 & 52,165 & 52,555 & 390\\
Soft 4 & 52,800 & 52,850 & 100\\
Soft 5 & 53,275 & 53,475 & 80\\
\enddata
\end{deluxetable}


We divide the ASM observations into periods of soft/high and consider
the remainder of the period to be hard/low (Table~1, Figure~1). There
are short intervals that appear similar to the soft/high states that we
do not single out, but treat as part of the remaining hard/low state. We
choose the time interval of the Soft~3 period to correspond with the
interval chosen by Lachowicz et al. (2006).

\begin{figure}
\includegraphics[width=1.0\textwidth]{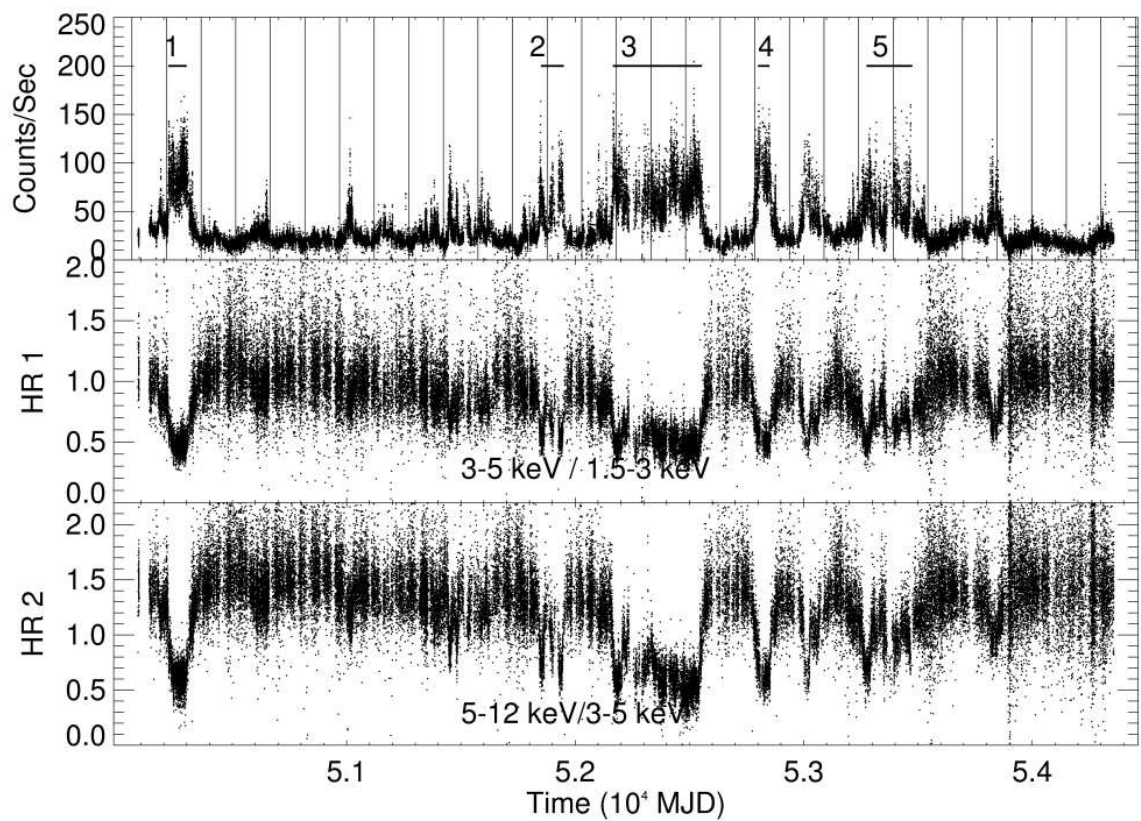}

\caption{The RXTE ASM light curves and hardness ratios over a 4270 day
interval from the start of the ASM observations. We show the light curve of
total counts, the hardness ratios HR~1 and HR~2, and the light curves
in the 1-3 keV, 3-5, and 5-12 keV ranges. Horizontal bars in top panels
show the definitions of the Soft 1, 2, 3, 4, and 5 states. Vertical bars in
the top left panel show times of superorbital phase 0 according to
the ephemeris of Lachowicz et al. 2006.}

\label{fig:lightcurves}
\end{figure}

\subsection{RXTE PCA/HEXTE Observations}

To supplement our analysis of RXTE ASM data, we also use pointed RXTE
PCA/HEXTE observations. We consider the data set of 202 observations
from 1999 through 2004 examined in 
Wilms et al. (2006).  These observations were scheduled biweekly for lengths of
$\sim5$~ksec. Wilms et al. fit the spectra between 3 and 200~keV to three
models: a broken power law with exponential cutoff, the Comptonization
model of Titarchuk (1994), and the Comptonization model of Coppi (1999).
These fits were used to explore the hard and soft states, and led to the
conclusion that there is a continuum between these states.
However, the fits have not previously been used to investigate the
5.6~day orbital period.

\section{Dip Variability}

\subsection{Soft State Periodicty with the ASM\label{sec:asm}}

We apply the Lomb Normalized Periodogram (Scargle 1982) and the Analysis
of Variance (ANOVA, Davies 1991) methods to search for periodicity in
the RXTE ASM observations of Cyg~X-1.

\begin{figure}
\includegraphics[width=1.0\textwidth]{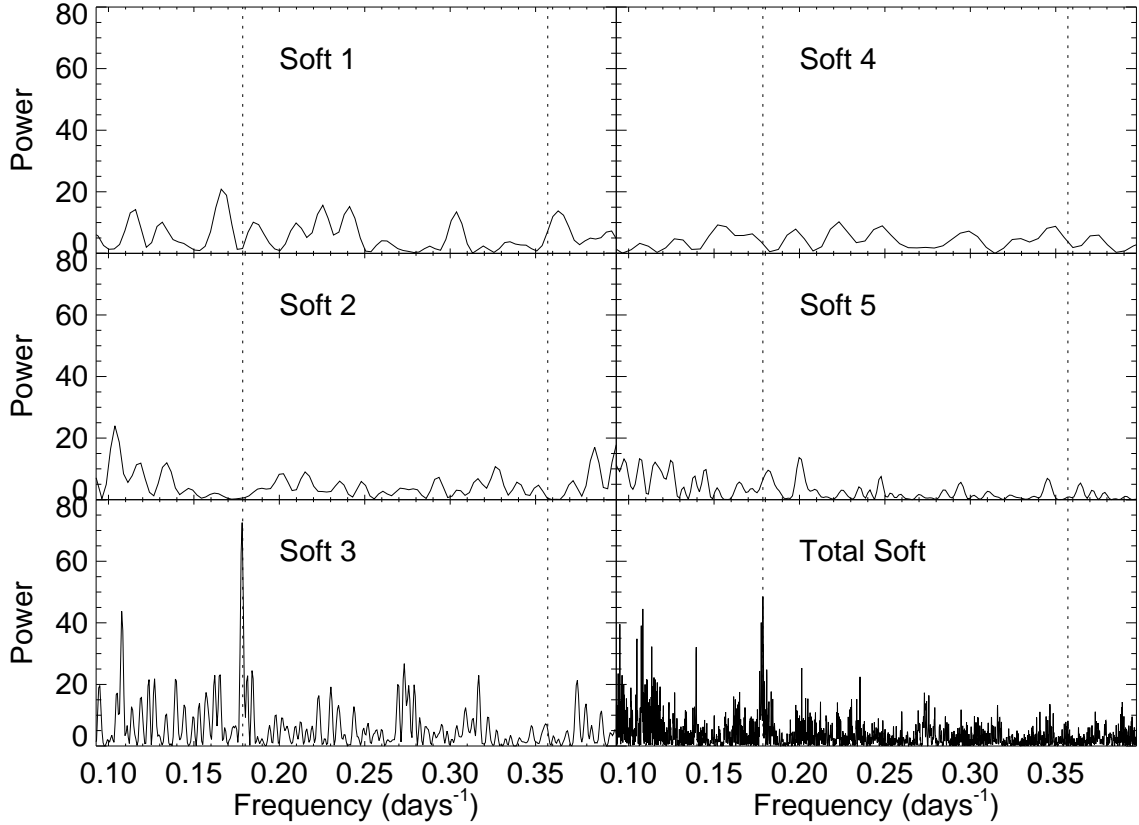}
\caption{The Lomb Normalized Periodogram of the H1 hardness ratio of
 Cygnus X-1, as observed by RXTE ASM during the 5 soft periods we have
selected. The frequencies corresponding to a 5.6 day orbital period and
its first harmonic are marked with vertical dotted lines.
\label{fig:lnp}}
\end{figure}

The Lomb Normalized Periodogram is a generalization of the Fourier power
spectrum to unevenly sampled data. In the case of evenly
sampled time series, the Lomb Normalized Periodogram is identical to the
power spectrum.
When we apply this periodogram to the hardness ratio HR~1 (as defined in
Figure~1) during the Soft~3 period, we find a peak power of 73 at a
period of 5.61 days and a power of 47 in the neighboring bin at 5.59
days. We eliminated points with HR~1 below 0 or above 3, as negative
hardness ratios are not physically meaningful and large hardness
ratios result from division by count rates near 0 and carry large 
uncertainties.
This only eliminated 2 of 6038 data points in the soft
state. We show the Lomb
Normalized Periodogram for each of the 5 soft states we have identified
in Figure~\ref{fig:lnp}.

For a power spectrum of uniformly sampled data, normalized so that the
white noise level is 1, the probability of a signal level $s$ would be
$\exp(-s)$. A LNP power of 73 corresponds therefore to a false alarm
probability of $2\times10^{-32}$. However, the signal at 5.6~days does
not appear against white noise, but instead against a continuum
strongest at low frequencies. Examination of the distribution of powers
shows that they obey a distribution closer to $\exp(-s/4)$, which would
imply a false alarm probability of only $\approx10^{-8}$, implying a
much lower significance of the signal.

\begin{figure}
\includegraphics[width=1.0\textwidth]{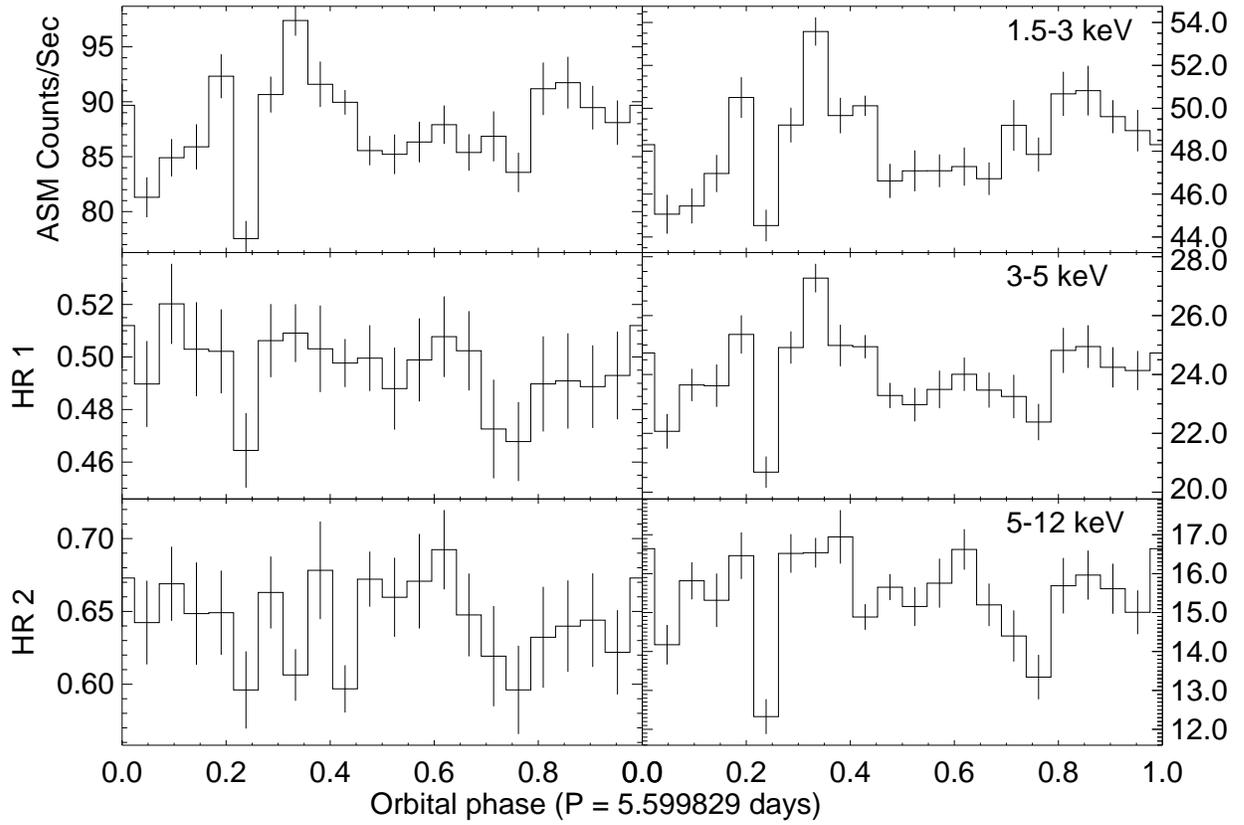}
\caption{Light curves and hardness ratios folded with the 5.6 day Cyg~X-1
 orbital period during the Soft 1 period.
\label{fig:folded}}
\end{figure}

\begin{figure}
\includegraphics[width=1.0\textwidth]{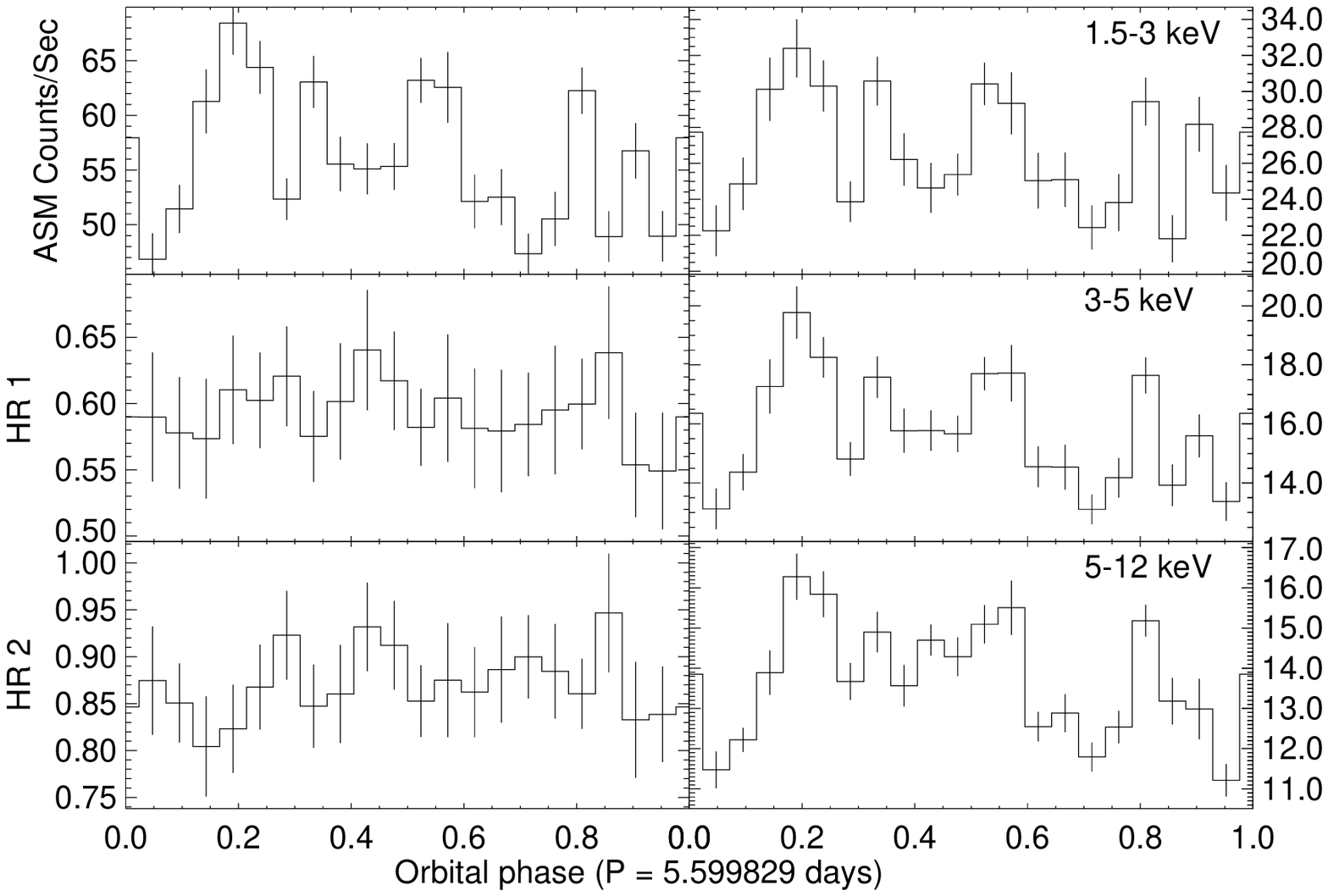}
\caption{As in Figure 3, for the Soft 2 period.}
\end{figure}

\begin{figure}
\includegraphics[width=1.0\textwidth]{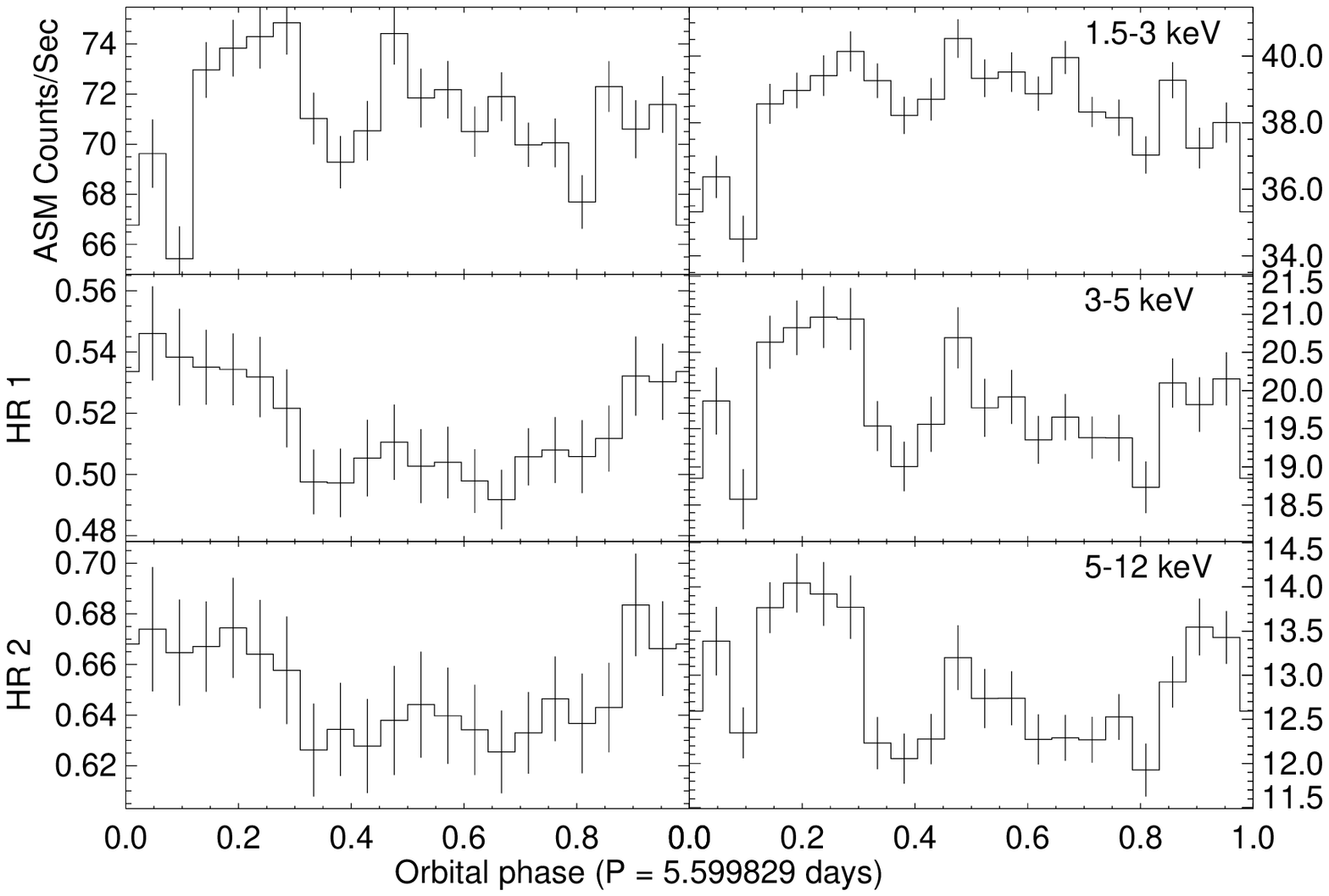}
\caption{
As in Figure 3, for the Soft 3 period.}
\end{figure}

\begin{figure}
\includegraphics[width=1.0\textwidth]{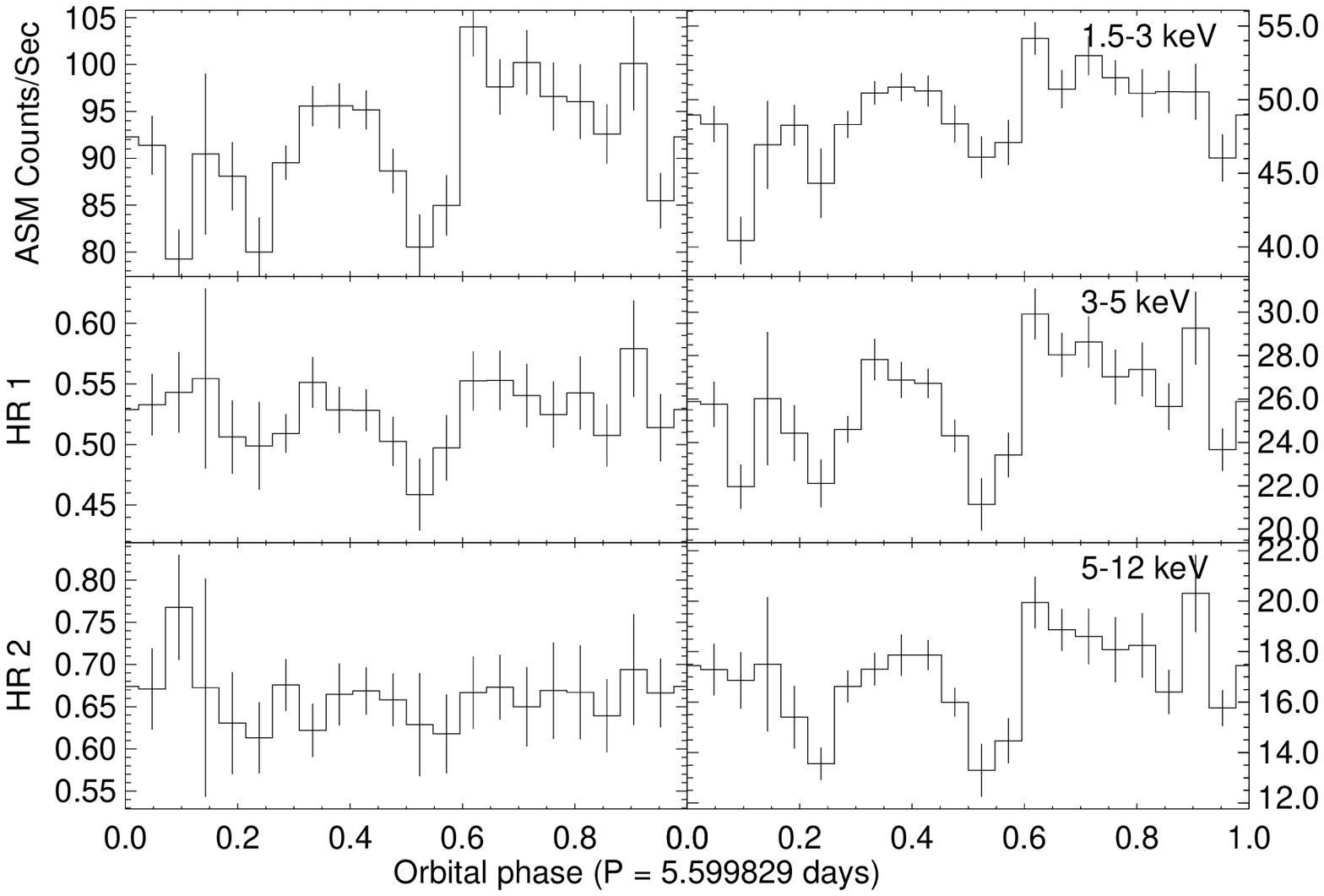}
\caption{As in Figure 3, for the Soft 4 period.}
\end{figure}

\begin{figure}
\includegraphics[width=1.0\textwidth]{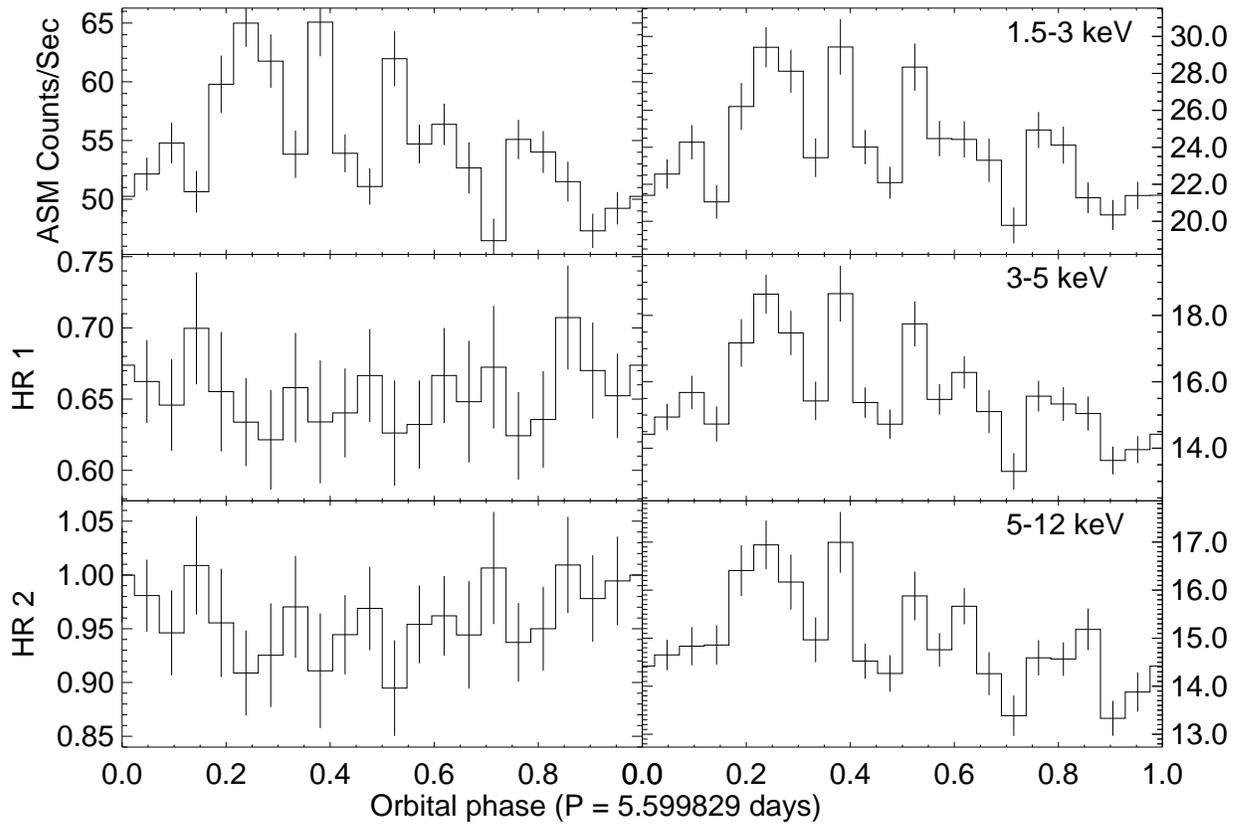}
\caption{As in Figure 3, for the Soft 5 period.}
\end{figure}

Application of ANOVA to the HR~1 ratio in the Soft~3 period gives
a probability of $5\times10^{-23}$ that variation at 5.6~days is random.
For this test, we used 16 phase bins. 
As with the power spectral methods, the ANOVA method may return a
false positive in a test for periodicity if the power spectrum
contains a significant continuum component, and the significance given
by this method may be an overestimate.

%
%

We have found a clear orbital variation in the hardness ratio HR~1,
but not in the individual energy bands in the same period (``Soft
3''). The count rates in the individual energy bands are more variable
on all timescales than HR1.  For example,
during Soft~3, the $\chi^2_\nu$ fit for a constant count rate in the
1.5-3~keV range is $62$ with 6037 degrees of freedom. The $\chi^2_\nu$
fit for a constant HR~1 is 7.1. As a result of this greater
non-orbital variability, it may be harder to detect any orbital
variation in the actual count rates than in HR1, which may increase
near $\phi=0$ with the absorbing column, but which may be less
affected by flares than the count rates.

If the Soft~3 orbital variation is $n \sigma$ significant,
then scaling from the duration of the observations, the Soft~1 and
Soft~2 variability should be $\approx 0.4 \sigma$ significant and
the Soft~5 variability should be $\approx 0.7 \sigma$
significant. However, the rms variability in HR1 is 50\%\ higher for
Soft~5, so that we might expect only $\approx0.5 \sigma$ significance
in the orbital variations.  The rms variation in the 3 X-ray bands and
2 hardness ratios are within a factor of 2 for all the Soft states,
except for HR2 in Soft~2, which appears anomalously high.

We also tested whether the HR1 variability seen during the Soft~3 period
would appear significant in shorter intervals of that same period.  
Lomb Normalized Periodograms of the first or middle 200 days of
Soft~3 show no significant variation at the orbital period, though the
power spectrum of the last 200 days shows a peak 
as great as for
the total Soft~3 period.  There was no significant variation in HR1 in
the first 100 days of Soft~3, or in days 190 through 290, but within
the last 100 days the power spectral peak reached 65. For the last 50
day interval in Soft~3, the power spectral peak is 47, but no
significant peak was seen for the first or second 50 day periods, or
another 50 day period near the end of Soft~3. Intervals of 200 days, 100
days, and 50 days are identical to the lengths of the Soft~1, Soft~4,
and Soft~2 periods, respectively. In conclusion, it is possible
that orbital variability as strong as that seen during the Soft~3
state could occur during all soft states but went undetected by RXTE/ASM
during the shorter Soft~1, Soft~2, Soft~4, and Soft~5 states.

In Figures~3 to 7, we present light curves and hardness ratios during the
Soft states 1 to 5, folded with the 5.6~day binary period.

\subsection{Individual dips seen with the RXTE ASM}

We have demonstrated with both power spectra and epoch folding methods that
the hardness ratio and count rate show the 5.6~day orbital
periodicity during the Soft~3 state.
However, previous studies of dips in the hard state have suggested
that discrete dips may occur
at any phase, but may simply be more likely to occur near $\phi=0$.

Cyg X-1 shows both dips in which the spectrum hardens (color dips) and
also monochromatic dips (count dips) in which the count rate diminishes but the
color does not change appreciably (Feng \&\ Cui 2002). We thus have 4
classes of dips to compare: color and count dips in the soft and hard
states. We choose criteria to define these dips.

\subsubsection{Color dips in the soft state}

Color dips in the hard state were already considered by
Baluci\'{n}ska-Church et al. (2000). They selected as dips those observations
with hardness ratios HR1$>2.0$ or HR2$>2.5$. We adjust these criteria to
the changed spectrum during the soft state.

\begin{figure} \includegraphics[width=8.2in]{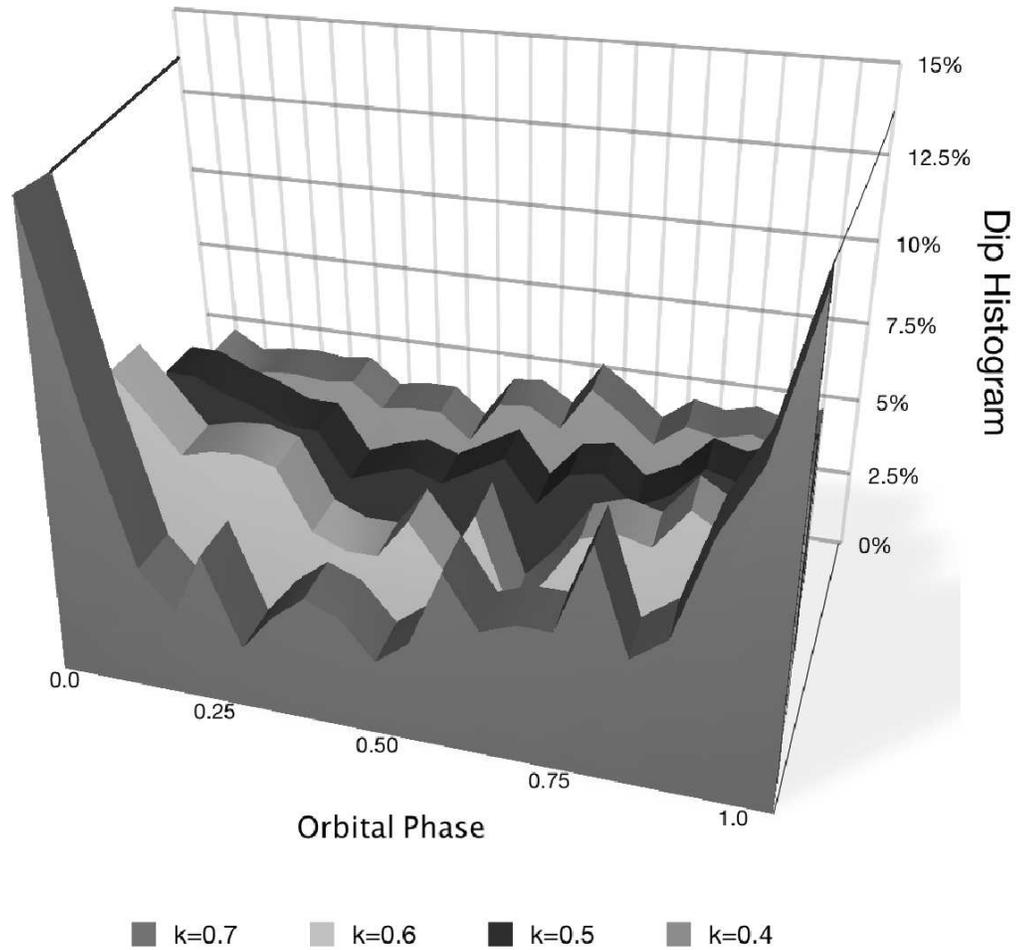}
\caption{\label{fig:histogramsoftcolor}Histograms of color dips versus orbital phase in the ``Soft~3''
state. The number of dips at each phase interval are divided by the
total number of dips. There are 20 phase intervals from 0.0 to 1.0. From
bottom to top, we show histograms of dips defined by
$\mbox{HR1}/\overline{\mbox{HR1}}>1.96 k$,  where
$\overline{\mbox{HR1}}$ is the mean
hardness ratio in the Soft~3 state.} 
\end{figure}

If we choose as dips those data points with
$\mbox{HR1}/\overline{\mbox{HR1}}>1.96$, where $\overline{\mbox{HR1}}$ is
the mean hardness ratio, we find only
14 observations to count as dips in the Soft~3 state.  From
\S\ref{sec:asm} we know that HR1 varies with the orbital period,
so a less restrictive criterion
should select individual observations (dips) with some preference for
orbital phase.  We thus choose some constant
$k<1$ and select dips based on $\mbox{HR1}/\overline{\mbox{HR1}}>1.96
k$.  The results, shown in
Figure~\ref{fig:histogramsoftcolor}, demonstrate that the strongest
dips defined by HR1 color cluster near $\phi=0$.  Weaker dips depend
less strongly on phase.

\begin{figure}
\includegraphics[width=8.2in]{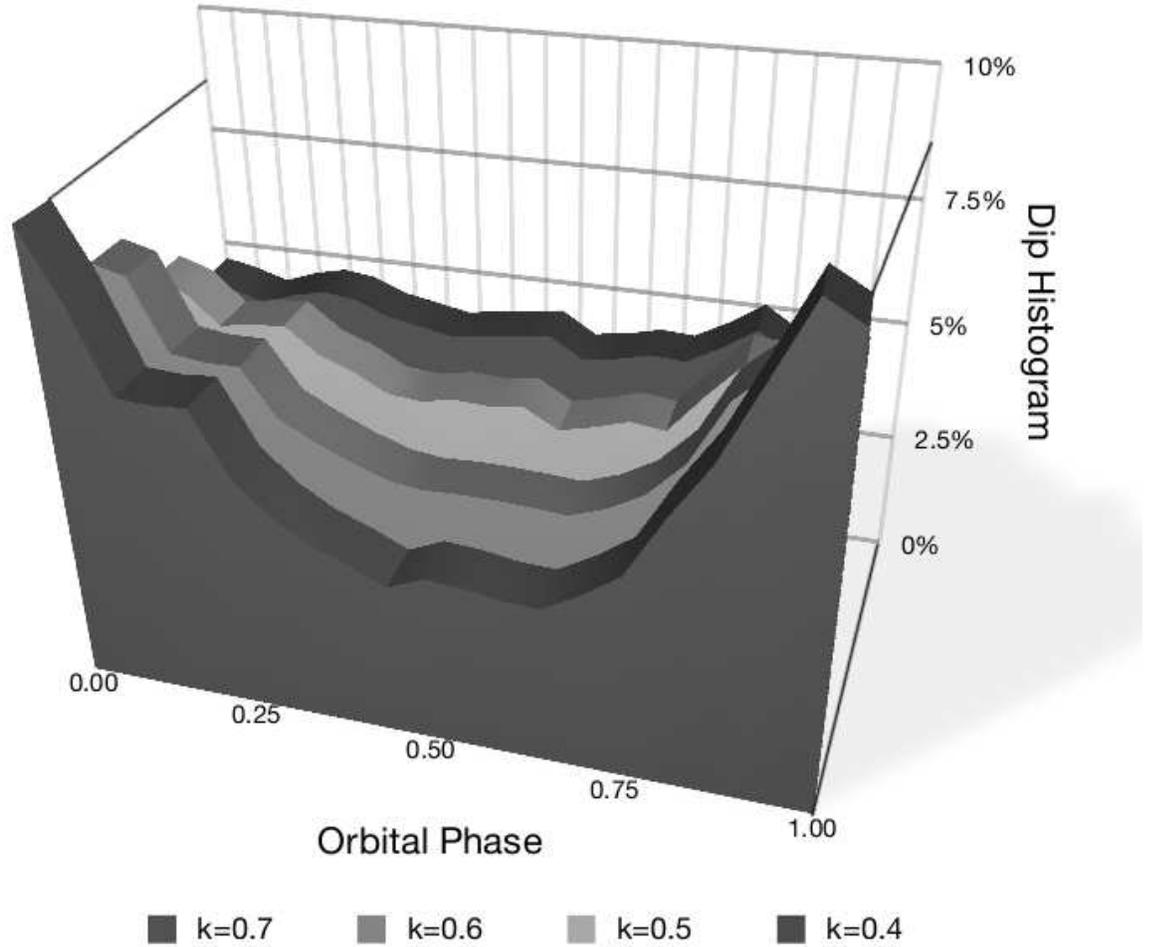}
\caption{\label{fig:histogramhardcolor}
Histograms of color dips versus orbital phase in 
the Hard state, for different depths of the dips. The number of dips at
each phase interval are divided by the total number of dips. 
We show stacked histograms of dips defined by
$\mbox{HR1}/\overline{\mbox{HR1}} >1.96 k$,
where $\overline{\mbox{HR1}}$ is the mean hardness ratio in the Hard state.}
\end{figure}

For comparison, we show a histogram of color dips in the hard state in
Figure~\ref{fig:histogramhardcolor}. As there is more data for the hard state than for the
Soft~3 state, this plot has less scatter. It shows more clearly the trend that stronger
dips are more likely to occur near $\phi=0$.

\subsubsection{Count dips in the hard state}

\begin{figure} 

\includegraphics[width=8.2in]{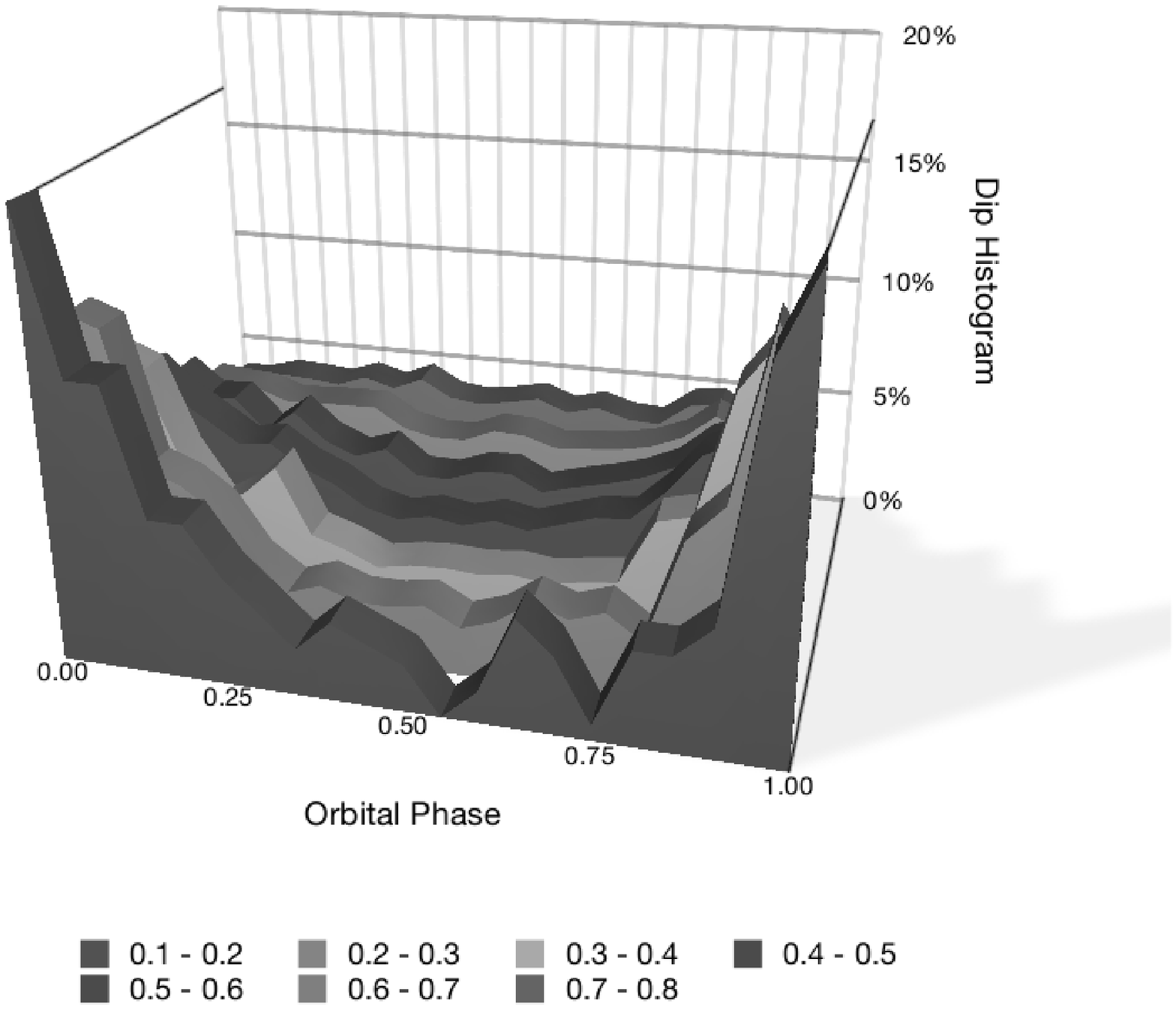}
\caption{\label{fig:histogramhardcounts}
Histograms of count dips versus orbital phase in the RXTE ASM
observations of Cygnus X-1 in the hard state.  The number of dips at
each phase interval are divided by the total number of dips. From bottom
to top, we show histograms of dips defined by
$0.1<$CR$/\overline{\mbox{CR}}<0.2$, $0.2<$CR$/\overline{\mbox{CR}}<0.3$, etc. 
up to $0.7<$CR$/\overline{\mbox{CR}}<0.8$, where CR is the count rate and
$\overline{\mbox{CR}}$ is the mean count rate in the hard state.}
\end{figure}

Dips selected by low count rate
(count dips) are less frequent in the soft state and show no obvious
orbital clustering. Count dips in the hard state cluster with phase
similarly
to color dips (Figure~\ref{fig:histogramhardcounts}).  To select 
dips based on count rate, we choose an interval for the constant $b$ and
select ASM observations
with count rate $\mbox{CR}=b \overline{\mbox{CR}}$.
We find that stronger count-selected dips are more
likely to occur near $\phi=0$.

\subsection{Hard State Variability with the RXTE PCA/HEXTE\label{sec:pointed}}

While the 202 pointed RXTE PCA/HEXTE observations fit to models by Wilms et al.
(2006) cover the Soft~3 period, they are not extensive enough to
show the orbital variability in the soft state.

However, the pointed observations suffice to show the orbital
variability in the hard state, already demonstrated with RXTE ASM observations.
The broken power law fits of Wilms et al., include the 
hydrogen column
density $N_{\rm H}$ as a free parameter.
In Figure~4 we show the fitted value of this parameter during the hard
state folded over the 5.6 day period, where we have defined the hard
state by $F(\mbox{5-10})/F(\mbox{2-5})>0.3$, where $F(\mbox{5-10})$ is
the 5-10 keV flux according to the model and $F(\mbox{2-5})$ is the 2-5
keV flux. Eliminating observations that have $N_{\rm H}=0$ leaves 133 data
points. Applying ANOVA, we find a false alarm probability
$4\times10^{-8}$ for a periodicity at 5.6~days.
If instead we examine $F(\mbox{5-10})/F(\mbox{2-5})$ during the hard state,
we have 145 data points which give a false alarm probability
of $8\times10^{-6}$.

Wilms et al. also fit the pointed RXTE observations to two Comptonization
models, one based on Titarchuk (1994) and another based on Coppi (1999).
The orbital variations in $N_{\rm H}$ and $F(\mbox{5-10})/F(\mbox{2-5})$ are not
as significant using fits with either of these models compared with
the fits using the broken power law model.


\begin{figure}
\includegraphics[width=1.0\textwidth]{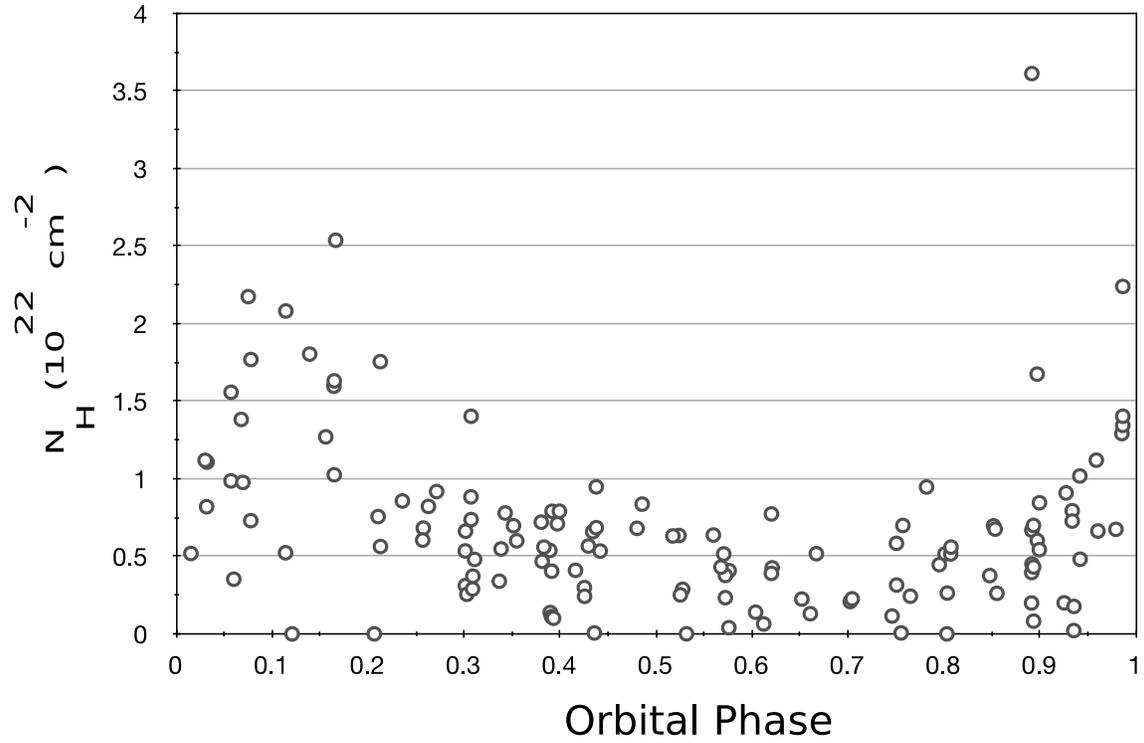}

\caption{We plot versus orbital phase the values of $N_{\rm H}$, the
hydrogen column density,
from fits of broken power law models to RXTE pointed observations of
Cygnus X-1 detailed in Wilms et al. 2006}

\end{figure}

\begin{figure}
\includegraphics[width=1.0\textwidth]{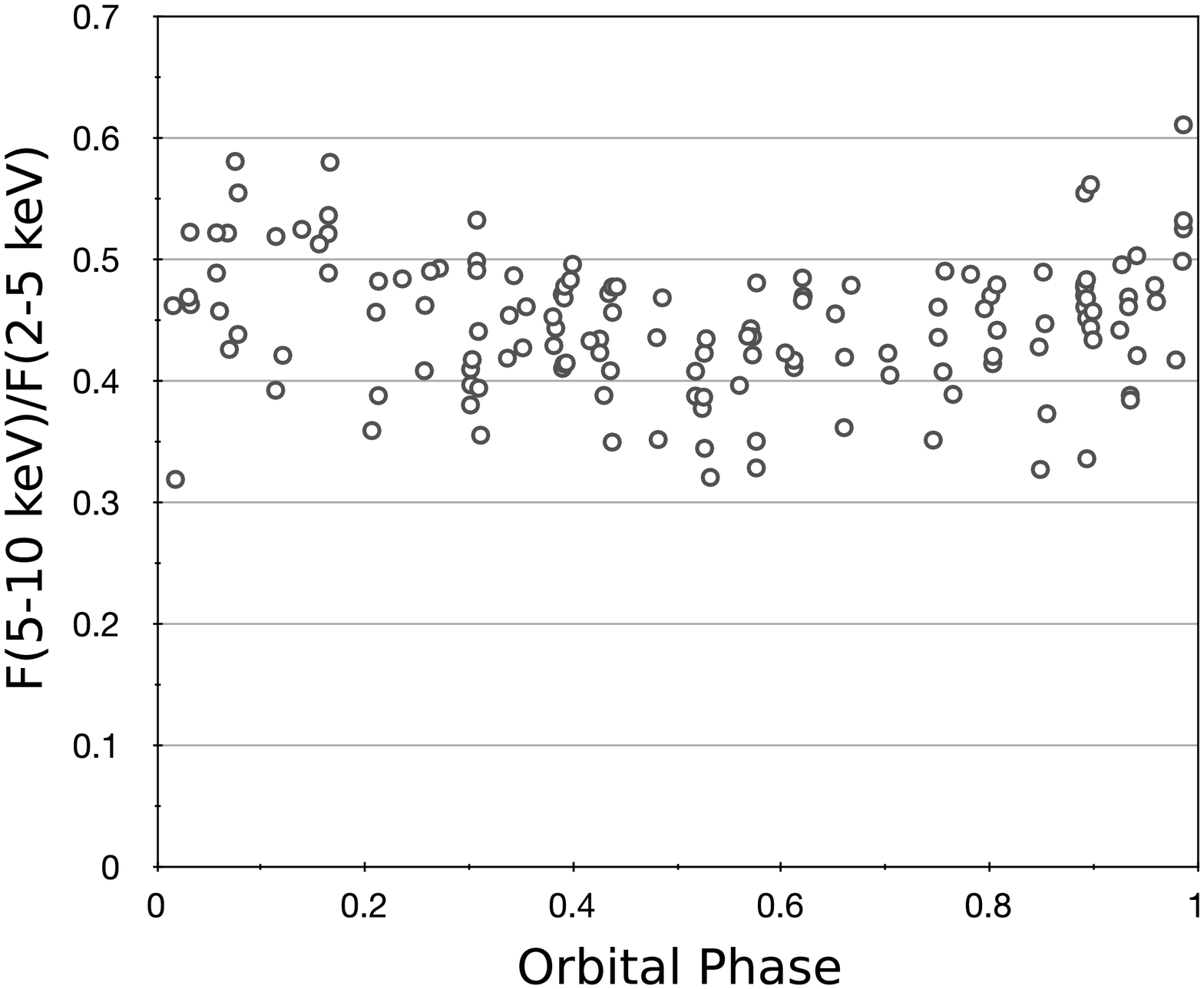}

\caption{Phase variability of $F(5-10)/F(2-5)$, the ratio of fluxes in
the 5--10 keV band to the 2--5 keV band, from fits to the pointed RXTE
observations presented by Wilms et al. 2006. The fits assume a broken
power law model.}

\end{figure}

The RXTE pointed observations have much higher spectral resolution,
timing resolution, and sensitivity compared with the ASM observations.
Here we have shown the orbital variation in absorbing column using not
just broad-band colors, but detailed fits to spectra.

\section{X-ray Absorption Models\label{xraymodel}}

From the increased X-ray absorption near $\phi=0$ during both the
soft states (\S\ref{sec:asm}) and hard states (\S\ref{sec:pointed}),
we can test models of how the stellar wind may change, if at all,
between those states.

First we assume that the stellar wind is identical during the soft and
hard states, and that the only change is in the X-ray spectrum ionizing
the wind. We compute the ionization in the stellar wind in response to
X-ray illumination by the soft or hard state X-ray spectrum. Then we
examine the limits
of wind mass loss rate and terminal velocity that are compatible with
orbital variability in each state.

\begin{deluxetable}{lccrrrr}
\tablecolumns{7}
\tablewidth{0pc}
\tablecaption{Model fits to the Cyg X-1 variability in three RXTE ASM
energy bands or to the hardness ratio between the two low energy bands. We give
here the X-ray luminosity $L_x$ and wind mass-loss rate $\dot{M}$ of the best
fit.}

\tablehead{
\colhead{State} & \colhead{Data} & \colhead{Model}
& \colhead{$\chi^2_\nu$} & 
\colhead{$L_x/10^{37}$} &
\colhead{$\dot{M}\times10^{6}$} 
& \colhead{$F_c$}\tablenotemark{a} \\
\colhead{}      &  \colhead{}    &  \colhead{}     &   \colhead{}
& 
\colhead{(erg s$^{-1}$)} & 
 	\colhead{(M$_\odot$~yr$^{-1}$)} & \colhead{}\\}
\startdata
Soft & 1.5--3, 3--5, 5--10 keV & Smooth wind & 3.5 & 0.25 & 0.36 & NA \\
Soft & 1.5--3, 3--5, 5--10 keV & Partial covering & 3.3 & NA & NA & 0.14 \\
Soft & 3--5 keV / 1.5--3 keV & Smooth wind & 1.10 & 0.25 & 0.72 & NA \\
Soft & 3--5 keV / 1.5--3 keV & Partial covering & 0.9 & NA & NA & 0.28 \\
Hard & 1.5--3, 3--5, 5--10 keV & Smooth wind & 3.2 & 1.7 & 1.8 & NA \\
Hard & 1.5--3, 3--5, 5--10 keV & Partial covering & 3.1 & NA & NA & 0.30 \\
Hard & 3--5 keV / 1.5--3 keV & Smooth wind & 2.8 & 1.5 & 2.2 & NA \\
Hard & 3--5 keV / 1.5--3 keV & Partial covering & 2.6 & NA & NA & 0.30 \\
\enddata
\tablenotetext{a}{Covering fraction}
\end{deluxetable}

\subsection{Input X-ray Spectrum\label{sec:inputxray}}

The ionization in the stellar wind depends on the X-ray spectrum. To
compute the ionization in the wind, we use the XSTAR program (Bautista
\&\ Kallman, 2001).
Under the assumption that the wind is optically thin to the X-rays from
the vicinity of the black hole, the ionization and temperature
equilibria can be computed based on the ionization parameter $\xi=L/n
r^2$ where $L$ is the X-ray luminosity, $n$ is the hydrogen density, and
$r$ is the distance to the X-ray source.

XSTAR requires as input the luminosity $L$ in the 13.6 eV to 13.6 keV
range.  Extending the power law to low energies, particularly during the
soft state, would lead to an unrealistic spectrum. We model the spectrum
using the Bulk Motion Comptonization (BMC) model (Shaposhnikov \&\
Titarchuk 2007), which has been implemented in the XSPEC spectral
fitting program. This model provides a physically-motivated spectrum
that is similar to a blackbody and a broken power law.

However, there may be an additional soft blackbody component with
$kT\sim0.1-0.2$~keV, as observed by Ebisawa et al. (1996) with ASCA.
 
We therefore simulate the orbital variability in X-ray absorption using
spectra given by the BMC fits of Shaposhnikov \&\ Titarchuk (2007) with
or without an added soft blackbody component of $kT=0.1$~keV. We
present here only the results without the soft blackbody, which has a
greater effect on the UV P~Cygni lines than on the X-ray absorption. For each
model we identify both the 13.6 eV to 13.6~keV luminosity, which is what
XSTAR requires, and the 2-200~keV X-ray luminosity, which describes the
X-ray output in the range observed by RXTE.

\subsection{Smooth, Partially Ionized Wind Absorption Model}

We make a model similar to that of Wen et al. (1999), in which
the orbital variation of X-rays results from the black hole
passing behind larger column densities of partially ionized wind near
$\phi=0$.

We fix the orbital inclination to 30 degrees, the radius of the O star
to $R_O=$1.387$\times10^{12}$~cm, and the orbital separation to 2~$R_O$.
We fix the radial velocity law in the wind to be $v(R)=1430
(1-R_O/R)^\beta$ with $\beta=0.75$. The wind parameters are based on the
fits to the UV P~Cygni lines observed during the X-ray soft state with
the HST {\it STIS}, as reported
in Vrtilek et al. (2008).

We calculate, in phase intervals of $\Delta \phi=0.05$, the column density
of the sight line to the black hole at each ionization parameter between
$\log \xi=1$ and $\log \xi=5$. This quantity, $N(\phi,\log \xi)$, can be
adjusted for various values of the wind mass loss rate $\dot{M}$ and the
X-ray luminosity $L_x$, by 
\begin{equation}
N^\prime(\phi,\log \xi)=\frac{\dot{M}^\prime}{\dot{M}} 
N(\phi, \log \frac{L_x \dot{M}^\prime}{L_x^\prime \dot{M}} \xi )
\end{equation}
where $\dot{M}$ and $L_x$ have been changed to take on values
$\dot{M}^\prime$ and $L_x^\prime$ respectively.

From XSTAR, we calculate the transmission of the source spectrum as it
passes through each column at each value of $\log \xi$, and thus the
X-ray spectrum at each phase $\phi$.  We add the Compton scattering
column density to the photoabsorption column predicted by XSTAR.
This allows us to find those values of $\dot{M}$ and $L_x$ that predict
the orbital variation in the Cyg~X-1 X-ray spectrum that results entirely
from the black hole's motion through a smooth, partially ionized stellar
wind.

The results of fitting the model to the orbital variation are shown in
Table~2. The best-fit values provided for $\dot{M}$ and $L_x$ are
uncertain, as these parameters vary in tandem.
The contours of reduced $\chi^2$ provide a
measure of which ranges of $\dot{M}$ and $L_x$ are acceptable,
although the models only provide fits with $\chi^2_\nu\approx1$ for
HR1 during the Soft~3 state. In the hard state, the models 
strongly favor $L_{\rm x, 37}/\dot{M}_{-6}<1$, $\dot{M}_{-6}<4$, while
in the soft state, they favor $L_{\rm x, 37}/\dot{M}_{-6}<1.5$
and $\dot{M}_{-6}<2$, where we use $L_{\rm x, 37}\equiv L_{\rm x}\times
10^{-37}$~erg~s$^{-1}$ and $\dot{M}_{-6}\equiv \dot{M}\times 10^{6}
{M_{\odot}}$~yr$^{-1}$.  For the soft state, the $\chi^2$ contours given by
fits to the HR1 color and to the individual X-ray bands are
complementary, limiting $L_x/\dot{M}$ and $\dot{M}$, respectively.

\subsection{Cold Wind Clump Absorption Model}

The discrete X-ray dips seen in Cyg~X-1, particularly near $\phi=0$
(Baluci\'{n}ska-Church et al. 2000), may be related to the dips seen in
such X-ray binaries as Hercules X-1 (e.g. Shakura,
Prokhorov, Postnov, \&\ Ketsaris 1999), or the class known as
``dippers'' (Frank, King, \&\ Lasota 1987). In these systems, dips
may be associated with the gas stream feeding the accretion disk
through Roche lobe overflow.
The focused wind in Cyg~X-1 could provide a similar absorbing structure.
However, a full treatment of a focused wind would be beyond the scope
of this paper, and would probably require hydrodynamic modeling.

Instead of dealing with complex asphericity, we pursue an alternative
to the smooth, partially ionized wind model in which the wind is
still assumed to be spherically symmetric and the orbital variation in
the X-ray spectrum still results from the line of sight at
$\phi=0$ passing through more of the dense wind regions. However, we
assume that the absorption is caused by neutral clumps, and that the
chance of such a clump passing through the line of sight is linearly
proportional to the wind column density. We fix the clump column density
to $N_H=4\times10^{22}$~cm$^{-2}$, following the ASCA observations of
X-ray dips (Ebisawa et al. 1996). We allow two free parameters, the
covering fraction of the clump when the wind column density reaches its
maximum, and the mass-loss rate in the wind. The latter influences the
X-rays only through Compton scattering, as the wind
outside of the dense clumps is entirely ionized. The results do not
constrain $\dot{M}$ very well, although at too high values, too great an
orbital variation is seen.  The covering fraction parameter
serves as a time-average of the actual covering fraction during dips 
with the zero covering fraction in periods when dips are not present.

\subsubsection{Hard State}

We first fit the model to the ASM light curves in the 1.5--3, 3--5, and
5--10 keV bands during the hard state. We define the hard state as the
entire light curve excluding the soft states identified in Table~1. This
may include some ``failed transistions'' to the soft state, or
intermediate states.

The best-fit values for the smooth wind model are
$\dot{M}\approx2\times10^{-6}$~M$_{\odot}$~yr$^{-1}$ and
$L_x\approx2\times10^{37}$~erg~s$^{-1}$. For the computation of
$\chi^2_\nu$, we use errors for the binned light curve based on the
variance within each bin.

We also try fits to the hardness ratio HR1 (flux in the 3--5 keV band
divided by flux in the 1.5--3 keV band). 


For the hard state, the dense clump model using partial covering
and the smooth partially ionized wind model fit the data equally well.

\subsubsection{Soft State}

We apply again the model of absorption in a partially ionized smooth
wind but now to the ``Soft~3'' period. 
The strong orbital variation in HR1 is better explained by a neutral
absorber than by a smooth partially ionized wind.
The values of $\dot{M}$ and $L_x$ which provide the best fit to the
variations in 1.5--3, 3--5, and 5--10~keV bands predict a 3.6\%\ 
peak to peak variation in HR1, whereas an 11\%\ variation is actually
observed.  A neutral absorber would absorb most strongly
the soft X-rays, thus increasing the variation in HR1.

For the neutral clump model, the
free parameters are the covering fraction and $\dot{M}$ in the fully
ionized remainder of the wind, which affects the fit only
through the variable column for Compton scattering.  For the 1.5--3,
3--5, and 5--10 keV bands, we obtain $\chi^2_\nu=3.3$ at a covering
fraction of 0.14, while for HR1 we obtain $\chi^2_\nu=0.9$ at a covering
fraction of 0.28.  The larger covering
fraction required to fit the hardness ratio is consistent with the
hardness ratio giving a wider variation than expected from the variation
in the individual bands. That the covering fractions differ in the
fits to the individual energy bands and to the hardness ratio may
indicate a deficiency in the model. The $\chi^2_\nu$ values for the partial
covering neutral clump model are comparable but slightly superior to the
$\chi^2_\nu$ values for the smooth partially ionized wind absorption
model.

\section{Discussion}

We are able to set some broad limits on how the mass-loss rate and
X-ray luminosity may vary between the hard and soft states. Consistent
with the earlier results of Gies et al. (2003), the limit that we
have set on the wind mass-loss rate in the Soft state
($\dot{M}_{-6}<2$) is lower than the limit on the mass-loss rate in
the Hard state ($\dot{M}_{-6}<4$). However, the models make many
simplifications and the wind in Cyg~X-1 is probably complex and aspherical.

The X-ray dips seen in Cyg~X-1 may be caused by high density
regions of the wind passing through the line of sight. The wind density
could be increased locally from some interaction between the wind and
the compact object or globally from instabilities common to the radiative
acceleration of O~star winds. Density contrasts of $\sim10^3$ are
commonly seen in simulations of O~star wind acceleration (Owocki,
Castor, \&\ Rybicki 1988). From the duration of the observed dips,
$\Delta \phi \lae 0.005\sim 200 \mbox{s}$, the clumps would have linear
size $\sim 10^{10}$~cm, given that the compact object moves with
velocity $360$~km~s$^{-1}$ and wind velocities are typically $\sim
1000$~km~s$^{-1}$. 

For a density contrast of 10$^3$, a typical spherical region of radius
$10^{11}$~cm would contain one absorbing clump of radius $10^{10}$~cm.
We assume that the wind outside of the clump does not contribute to
absorbing dips. There would be $\sim10$ absorbing clumps within a
stellar radius, each clump presenting $\sim0.01$ of the surface area of
the surrounding $10^{11}$~cm radius sphere. Thus we would expect a clump
covering fraction of $\sim0.1$, which is consistent with
the results of the simple models shown in Table~2. 

It has been suggested (Kaper et al. 1993) that the flares seen in some
accreting binaries result from accretion of inhomogeneities in the
stellar wind.  If, as suggested above, a typical clump has radius
10$^{10}$~cm and is overdense by a factor of $\sim10^3$, then it will
provide a mass of $\sim10^{20}$~g. With an accretion efficiency of
0.1, this would provide a luminosity of $\sim10^{38}$~erg~s$^{-1}$
over a period of 100~s, consistent with observed flares. However, this
analysis does not explain the dependence of flaring on X-ray state.
Our suggestion that the wind mass-loss rate $\dot{M}$ is
greater during the hard-low state appears to be inconsistent with
increased flaring during the soft-high state. Ducci et al. (2009)
modeled the accretion of clumps in the winds of supergiant High Mass
X-ray binaries and showed that a greater $\dot{M}$ should lead to
brighter flares. However, the flares also depend on the distribution
of clump size, as well as other wind parameters that may change
between the two states.

If a focused wind is present and resolved into clumps with the
parameters we have described, there may be observable implications for the
UV P~Cygni lines. The clump covering fraction for the O~star may 
be less than that for the black hole as a result of the orbital
inclination. Whereas the wind outside of the clumps will be mostly
ionized to stages higher than
\ion{N}{5} and \ion{Si}{4} by X-ray
illumination, we expect that most of the N in the clumps will be
at lower ionization stages than \ion{N}{5}, as the higher density will
reduce $\log \xi$ by 3. As a result, the clumps may not contribute to
the profiles of the \NV doublet. However, \ion{Si}{4} may be the dominant
ionization stage of Si in the clumps, and thus an absorption component of
depth $\lae5$\%, limited by the covering fraction, could persist at
$\phi=0.5$ in the \SiIV doublet. Such an absorption
feature varying between hard and soft states could not be ruled
out by present UV observations.

In some models of the Superfast X-ray
Transients (SFXTs, Sguera et al. 2005), a new class of X-ray source
discovered with INTEGRAL, accretion of clumps from an OB~star wind
play an important role (Walter \&\ Zurita Heras 2007). The 
study of absorption dips reported here complements the active research
on SFXTs, whose compact objects may accrete from stellar winds at
greater orbital distances.

\acknowledgements

Results provided by the ASM/RXTE teams at MIT and at the
RXTE SOF and GOF at NASA's GSFC.

This paper benefited from discussions with Jeff McClintock and Jan Vrtilek.

\clearpage

\end{document}